# Luminescent solar power – PV/thermal hybrid electricity generation for cost-effective dispatchable solar energy

*Shimry Haviv,[#1] Natali Revivo,[#1] Nimrod Kruger,[2] Assaf Manor,[1] Bagrat Khachatryan,[1] Michael Shustov,[1] and Carmel Rotschild[*1,2]*

[1]Faculty of Mechanical Engineering, Technion – Israel Institute of Technology, Israel

[2]The Nancy and Stephen Grand Technion Energy Program (GTEP), Technion – Israel Institute of Technology, Israel

KEYWORDS: high-temperature photoluminescence; solar energy; rare earth emitters; hybrid PV/thermal

ABSTRACT

The challenge in solar energy today is not the cost of photovoltaic (PV) electricity generation, already competing with fossil fuel prices, but rather utility-scale energy storage and flexibility in supply. Low-cost thermal energy storage (TES) exists but relies on expensive heat engines. Here, we introduce the concept of luminescent solar power (LSP), where sunlight is absorbed in a photoluminescent (PL) absorber, followed by red-shifted PL emission matched to an adjacent PV cell's band-edge. This way the PV cell operates nearly as efficiently as under direct illumination, but with minimal excessive heat. The PL-absorber temperature rises due to thermalization, allowing it to store the excessive heat, which can later be converted into electricity. Tailored luminescent materials that support an additional 1.5kWh PV-electricity for every 1 kWh of (virtual) heat engine-electricity, with a dynamic shift between the two sources are experimentally demonstrated. Such an ideal hybrid system may lead to a potential reduction in the cost of electricity for a base-load solution.

INTRODUCTION

Concentrated solar power (CSP) denotes the technology wherein a thermal absorber is heated by the concentrated sunlight, thus enabling thermal energy storage (TES) for dispatchable generation[1–3]. CSP electricity generation is considered expensive in terms of the levelized cost of electricity (LCOE), roughly 7¢/kWh nowadays, in regions with high solar flux[4,5], compared to 2¢/kWh[6] for Si-based photovoltaics (PVs) in similar regional conditions. Nevertheless, recent industrial progress in small solar fields with higher efficiency[7], and new heat engines[8,9], as well as high-temperature hydrolysis[10], puts CSP as a vital alternative to batteries for baseload solutions due to its low LCOE (1¢/kWh[11]) portion for TES. Projection of a base-load solar energy LCOE below 3¢/kWh implies that, potentially, half the US energy production may come from solar by 2050[12] and also comply with the European green deal roadmap[13,14].

To utilize both heat energy and PV electricity from solar radiation, we may, for instance, look to the concept of solar thermal PV (STPV)[15,16], where the energetic tail of thermal emission is harvested by a low band-gap solar cell. Another relatively new concept called thermally enhanced photoluminescence (TEPL)[17] involves the coupling of PL from a low band-gap absorber to a high band-gap PV cell by thermally induced blue-shift. Despite very high theoretical maximal efficiencies, these concepts require high operating temperatures of about 2000°C and 1000°C, respectively, which limit their practicality. Furthermore, they cannot be used for TES as both require maintaining the heat for maximal PV efficiency. Alternatively, hybrid concentrated photovoltaic/thermal (PV/T)[18–20] methods, where electricity generation by a PV cell is performed in parallel to the extraction of unused heat, thus allowing TES, can potentially meet the requirements for constructing the sought after renewable base-load solution. Current PV/T techniques such as spectral splitting, where part of the solar spectrum is channeled to the PV cell while the other is channeled to a heat cycle, however, fall short because they sacrifice heat utilization for PV efficiency or vice versa.

The concept we propose here can best be explained thermodynamically by introducing the ideal PV/T concept of PV heating. A solar cell, when conventionally operating at 20–30% efficiency, converts the residual 70–80% of the incident solar power into heat. Conceptually, if the solar cell would work efficiently at high temperatures, 500°C for example, the heat accumulated on the cell could be converted into electricity in parallel through a conventional heat engine as in CSP. This would result in an additional 20–30% efficiency. Unfortunately, such an ideal hybrid PV/CSP solution cannot be done with solar cells, as their efficiency inherently drops as the temperature rises due to a reduction in open circuit voltage[21]. Nonetheless, what cannot be done with electrons in PVs can be done optically by PL. We note that as far as we know, PL in the context of solar energy was only used in luminescent solar concentrators[22,23] for driving PV cells, without utilizing the excessive heat for electricity.

RESULTS AND DISCUSSION

The PL process is very similar to what occurs in PVs. It conventionally involves the absorption of energetic photons, followed by thermalization and emission of low energy red-shifted photons. The external quantum efficiency (EQE)— the ratio between emitted photons and absorbed photons —does not necessarily depend strongly on the temperature of the PL-absorber[24,25]. When the emission is tuned to fit a PV cell's band-edge, the PL-absorber retains each photon's extra heat, while the PV cell converts the emission to electricity with minimal wasted heat[26]. This concept, namely luminescent solar power (LSP), removes the thermal portion at the single-photon level and enables an optimal spatial separation of heat energy from the free-energy, which is then available to the PV cell for maximal concentrated PV/T performance.

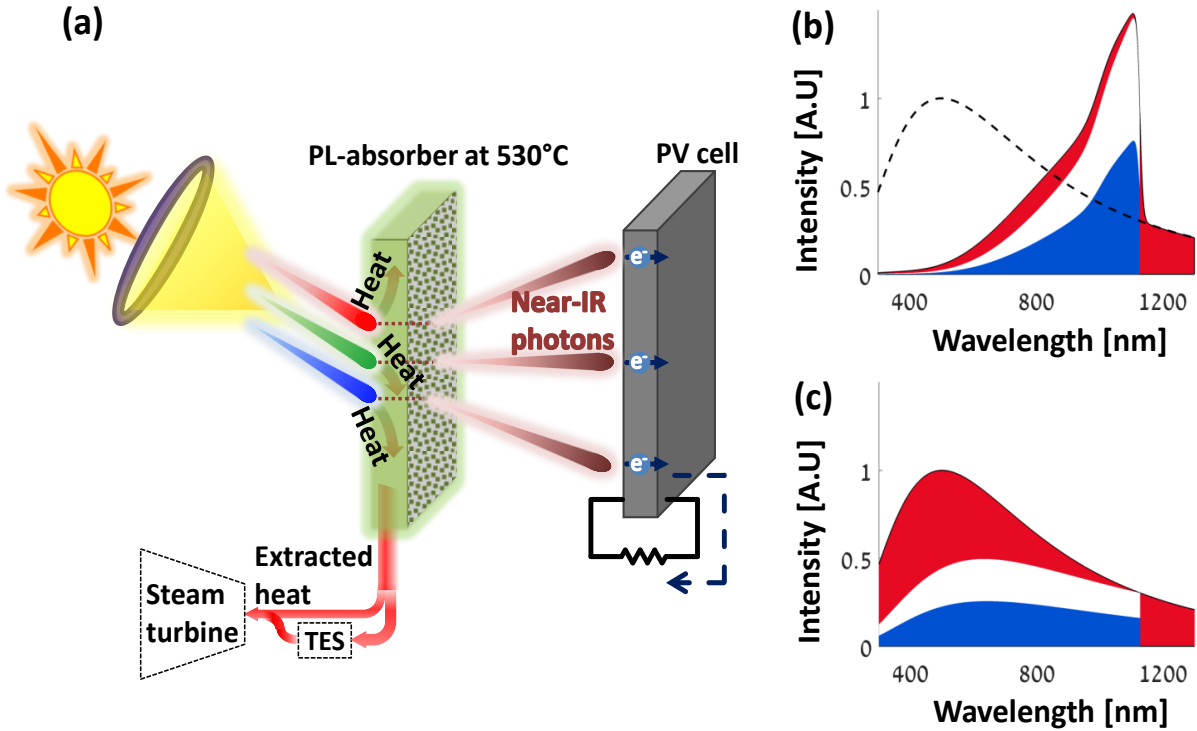

**Figure 1. Illustration of the LSP concept**. (**a**) Concentrated sunlight illuminates a PL-absorber. The red-shifted PL emission is then coupled to a PV cell with a matching band-gap, while the residual heat (at 530°C) is stored and transferred to a heat engine. (**b**) The PL's power spectrum is tailored to the PV cell's band-edge, compared to the solar spectrum (dotted line). The filled areas reflect the free energy utilized by the PV cell (blue), and the energy lost on thermalization (red). The white gap is associated with various losses, most of which also contribute to heat in non-ideal solar cells. (**c**) Similar representation of a PV cell under direct sunlight illumination.

Figure 1a pictorially details our concept. Solar radiation is concentrated on a PL-absorber that absorbs the high-energy photons and, after thermalization, emits low-energy photons with a high EQE at high temperatures. A diffusive surface at the back of the PL-absorber, together with a highly reflective coating at the front face directs the PL towards the PV cell's side. The emitted photons, combined with the transmitted ones, drive the adjacent PV cell, with a matching band-gap of 1.1eV, as nearly as efficiently as under direct illumination but without the excessive heat, which otherwise reduces the efficiency. This process by itself allows concentrated-PV (CPV) with minimal heating. More important, through the thermalization of the PL process, the PL-absorber accumulates the excessive heat and is heated to above 500°C. Heat transfer fluid (HTF) stores and later transfers the heat to a heat engine for electricity generation.

As an example of our concept, consider a single-junction InGaAs (or possibly InGaAsP) PV cell with a band-gap of 1.1 eV designed for CPV. A matching optimal PL-absorber material should have full solar absorption up to a desired cut-off wavelength and a narrow emission peak centered at a wavelength matching the PV cell's band-edge for minimal heating (Figure 1b). Much of the heat load induced by thermalization of energetic photons and solar infrared (IR) light, falls on the absorber material, leaving only residual heating at the PV cell (Figure 1b, red area). Comparing this to the heat load under direct illumination of an ideal PV cell (Figure 1c, red area) highlights the reduction in the required cooling. Remaining losses—angular mismatch, radiative, Boltzmann, and Carnot (all marked as white areas in Figure 1b and 1c)—also partially contribute to the heat load[26].

In addition, as will be shown later, by reducing the HTF flow-rate, the temperature of the PL-absorber increases, and the emission toward the PV cell is reduced. This way, LSP can shift more energy to storage or immediate supply (PV). Such flexibility in the power flow allows a twenty-four-hour "on-demand" supply with minimal waste, overcoming a major loss in PV/CSP systems, which have a predetermined capacity.

An efficient LSP device will depend on the optical performance of the absorber material, specifically, its solar absorption and emission spectra, and the EQE under the concentrated sun at temperatures relevant for TES. In the following, we demonstrate the above experimentally and analyze the PV cell's performance under the solar spectrum altered by the high-temperature PL emission.

For a 1.1eV PV cell's band-gap, ideal for the solar spectrum[27], the rare-earth (RE) elements neodymium ($Nd^{3+}$) and ytterbium ($Yb^{3+}$) with emission lines around 1μm are a perfect fit[28]. These materials also have a high EQE when doped in transparent matrices such as single-crystal (SC) and polycrystalline (ceramic) yttrium aluminum garnet (YAG), replacing different elements in the lattice[29,30]. The high EQE is retained at high temperatures due to the isolation of the electronic transitions from the phonons[25]. Common materials used for broad absorption and sensitization of $Nd^{3+}$, as used for flash-pumped lasers[31–33], are cerium ($Ce^{3+}$) and chromium ($Cr^{3+}$). We examined a variety of these dopant concentrations using both ceramic and SC models. The prominent comparison parameters are the EQE of the PL material and the full photon rate under broad sunlight excitation in the working temperature range. Of the variety we tested, a few stood out. We focus our study on four of them with weight percent (wt%) compositions of $A=\%Cr^{3+}, B=\%Nd^{3+}, C=\%Yb^{3+}$ doped in YAG: a ($A=0.5, B=1, C=1$) SC, and ($A=0.5, B=1, C=1$), ($A=3, B=1, C=1$) and ($A=3, B=1, C=3$) ceramics, namely, SC#1, Ceramic#2, Ceramic#3, and Ceramic#4, respectively. The samples typically have a broad absorption up to 650nm, and several narrow absorption lines in the near-IR (see Figure s1 in the SI). For each sample, we chose a thickness that absorbs about 60% of the total solar power at operating temperature (10mm for SC#1, 1.6mm for Ceramic#2, and 2mm for Ceramic#3 and Ceramic#4). In terms of the PV cell operation, 45% of the accessible solar photon rate at wavelengths shorter than 1130nm is absorbed. Figure 2a presents the measured spectral photon rate (number of photons per second per wavelength per area) of the solar radiation when no sample is placed (dashed black line). The solid lines present the measured spectral photon rate, of Ceramic#3 sample as an example, at elevated temperatures under full sunlight excitation (from blue at room temperature to red at 550°C). These results are typical for all the samples. Seen as an absent of photon rate from the solar spectrum, the samples absorb at wavelengths up to 830nm and emit between 830nm and 1130nm (observed as an enhanced photon rate above solar radiation). Our measurements show that an EQE of 92% at room temperature is reduced to 70% at 430°C, and drops to less than 45% at 530°C (see Figure s2b in the SI). The PL-absorber is mostly transparent for photons at wavelengths longer than 700nm, which have a lower thermal portion and may contribute less to heat. In the device schema, these photons reach the PV cell directly, maintaining its high efficiency even for a moderate EQE.

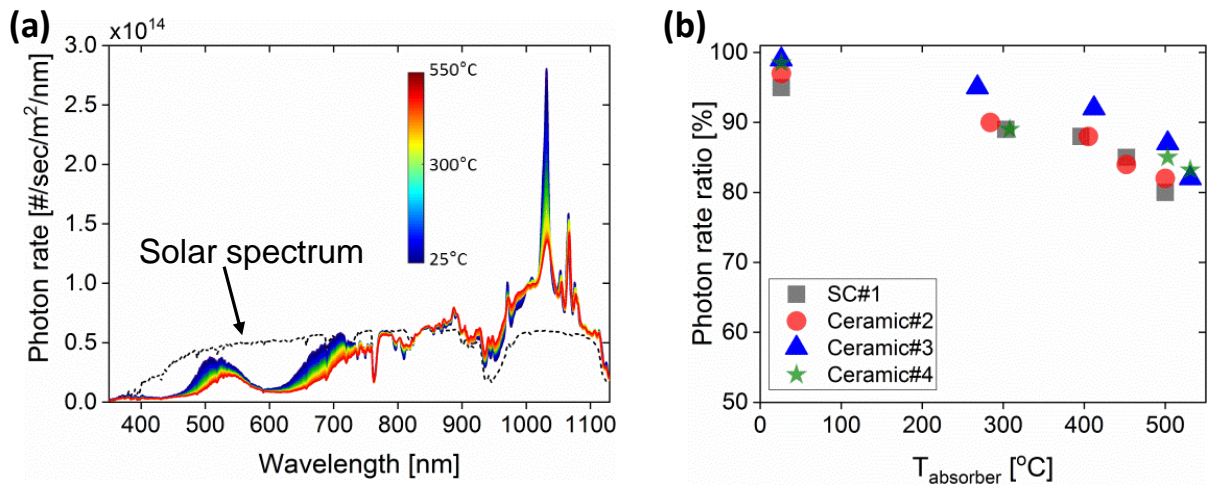

**Figure 2. Temperature dependent optical properties under solar excitation.** (**a**): Typical measured spectrum at elevated PL-absorber temperatures (solid blue line at room temperature, up to red at 550°C), compared to the measured direct illumination (black dashed line). (**b**) The ratio of the total photon rate in the wavelength range of 350–1130nm for each sample compared to direct illumination at various temperatures.

Figure 2b shows the ratio of the photon rate in the wavelength range of 350–1130nm available to the PV in the presence of the PL-absorber at various temperatures, compared to the measured solar photon rate. This ratio indicates the relative expected PV efficiency compared to its efficiency under direct solar illumination. As can be seen, a typical relative PV efficiency of above 95% at room temperature is reduced to 83% at 530°C. In a device, the temperature varies linearly along with the HTF flow under constant heat flux. The temperature difference between the outlet/inlet HTF temperatures depends on the inlet/outlet temperatures of the heat engine. Assuming HTF inlet/outlet temperatures of 280°C/480°C, with ΔT=50°C between the PL-absorber and the HTF, the average PL-absorber temperature is 430°C and the photon rate is expected to be 90% compared to direct illumination. Furthermore, the power spectrum indicates that the average photon energy impinging on the PV cell is reduced (red-shifted) from 1.9eV to 1.4eV. This reflects a reduction of the heat load of about 50%, assuming an operating voltage of 0.85V under a concentrated solar flux of 500–1000 suns (modified for similar parameters of GaAs from Ref. 34 and 35). A direct measurement of the heat load reduction and the PV cell operation, which supports these conclusions, is presented in the SI.

To expand our vision of LSP's potential performance, we simulated the device efficiency under real-world conditions. Figure 3a shows our system's two energy conversion mechanisms: the free energy part (PV electricity) and the thermal cycle part, which is also the storable part of the system. We modeled a PL-absorber with an EQE varying between 0 and 1, and full absorption up to a cut-off wavelength varying between 400nm and 1100nm. The emission peak of the PL was taken as 1064nm of the $Nd^{3+}$, matching a 1.1eV PV cell band-edge. The PV electricity portion, $PV_{portion}$, as a percentage of the total incoming power, was calculated using the following formula:

$$PV_{portion} = \eta_{PV} \left( R_{\#,abs} \cdot EQE \cdot \eta_{coupling} + (1 - R_{\#,abs}) \right) \quad (1)$$

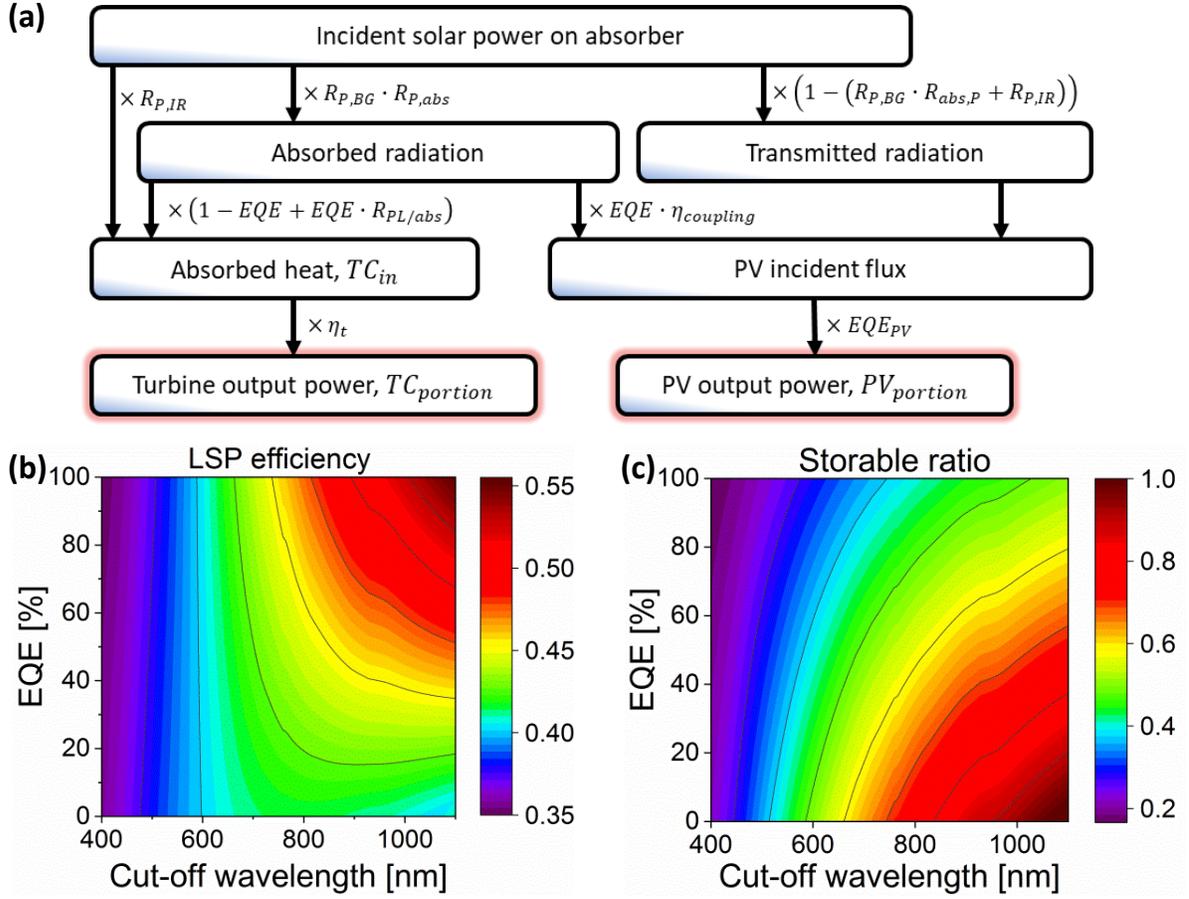

**Figure 3. LSP power flow.** (**a**) LSP power flow scheme. (**b**) The device-modelled total combined efficiency and (**c**) the portion of the thermal output power out of the total output power both as a function of the absorption cut-off wavelength and the average EQE of the PL-absorber.

where $\eta_{PV}$ is the PV cell's efficiency taken as 0.29[34,35] for the InGaAs PV cell under concentrated sunlight of 1000 suns and includes its quantum efficiency $EQE_{PV}$ as depicted in Figure 3a. $R_{\#,abs}$ is the ratio between the absorbed photon rate up to the cut-off wavelength and the photon rate up to 1100nm. $\eta_{coupling}$ is the extraction coupling efficiency between the PL-absorber and the PV cell, taken as 0.95 according to a ray-tracing calculation of various geometric shapes designed to enhance the PL toward the PV. The factor $1-R_{\#,abs}$ is the ratio of the transmitted photon rate, having a coupling coefficient to the PV cell of approximate unity.

The thermal cycle portion, $TC_{portion}$ was calculated using the following formula:

$$TC_{portion} = \eta_t \left( R_{P,BG} \cdot R_{P,abs} \cdot (1 - EQE + EQE \cdot R_{PL/abs}) + R_{P,IR} \right) \quad (2)$$

where $\eta_t$ is the thermal conversion efficiency including losses taken as 0.4 assuming outlet temperature of 480°C of the HTF[2]; $R_{P,BG}$ is the ratio between the solar power up to 1100nm and the entire solar spectrum taken as 0.83; $R_{P,abs}$ is the ratio between the absorbed solar power up to the cut-off wavelength and the total solar power up to 1100nm; $R_{PL/abs}$ is the ratio between the PL average photon energy and the absorbed average photon energy; and $R_{P,IR}$ is the ratio of the solar power at wavelengths longer than 1200nm and the entire spectrum, taken as 0.12, which can be absorbed in the front cover of the receiver using ITO nanoparticles[36,37] with matching resonance. The total combined efficiency of the system is calculated by $\eta_{tot} = PV_{portion} + TC_{portion}$ and presented in Figure 3b as a function of the cut-off wavelength and EQE.

In the LSP system, there is inherent mutual compensation between the PV portion and the thermal portion, which reduces losses compared to each conversion mechanism separately. It can be seen from Figure 3b that high performance of above 40% solar to electricity efficiency is achievable at broad working conditions (green-red portion). Figure 3c presents the ratio of the thermal output power, which can be stored, to the total output power.

Calculating the expected performances using our measured data of a-680nm cut-off wavelength and 70% EQE at an average temperature of 430°C, the result is that 45% of the power goes to the thermal part, thereby generating 17.8% electricity through the turbine. Another 24.4% is delivered by the PV, which results in an overall efficiency of 42.2%.

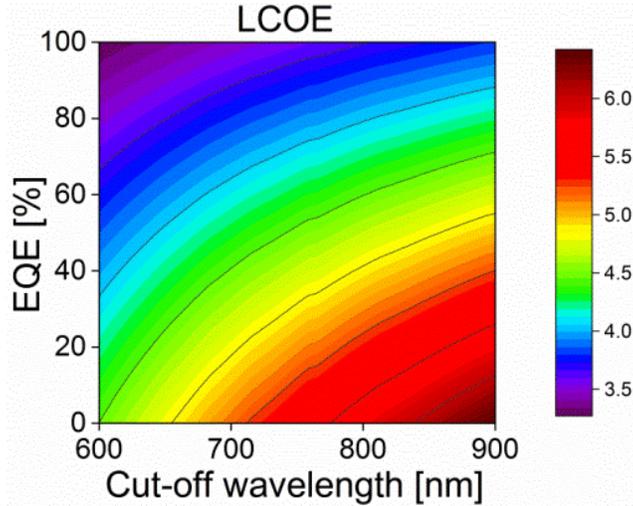

**Figure 4.** LSP LCOE estimation as a function of the absorption cut-off wavelength and the average EQE of the PL-absorber.

An additional feature of the LSP device comes from the photon rate reduction as the temperature rises, as evident in Figure 2b. This feature can be used for the "on-demand" shift of energy from the PV to the thermal storage, without an overall loss. This situation contrasts sharply with the conventional side-by-side PV-CSP, whose capacity is predetermined, leading to overcapacity and inherent losses. Increasing the average temperature from 430°C to 550°C reduces the photon rate from 90% to 75%, which adds 23% to the thermal portion. This temperature increase can be achieved by reducing the HTF flow, which reduces the heat convection coefficient. This, in turn, increases the PL-absorber temperature, which reduces the EQE and the emission toward the PV. The result is the increase of both the HTF's exit-temperature and the heat flow.

For a broader view of LSP, we also estimate its potential to reduce the LCOE. We note that LCOE estimation is challenging for any new concept. Here, we rely on the well-established LCOE of CSP with recent bids at 7¢/kWh LCOE for CSP[5] and an additional heliostat field at an LCOE of 1.7¢/kWh[38]. Additional costs of the PV and the PL-absorber add 0.23¢/kWh to our valuation[39], which is in agreement with the relatively small portion of the receiver's costs in CSP[34]. The total average LCOE can be estimated by dividing the cost of production by the total conversion output, formulated thus:

$$LCOE_{LSP} = [(7 + 0.23) * TC_{in} + 1.7 * (1 - TC_{in})] / \left[TC_{in} * \left(1 + \frac{PV_{portion}}{TC_{portion}}\right)\right] \quad (3)$$

where $TC_{in}$ is the percent of the absorbed heat from the total incoming power, which according to the experimental results is 0.45. The LCOE estimation is depicted in Figure 4 as a function of the cut-off wavelength and the EQE. As can be seen for the above experimental demonstration, an average LCOE of 3.72¢/kWh may be achieved. Any additional reduction in the CSP's LCOE as a result of a change in a component, such as a turbine, tower, and the solar field, will make a similar contribution to the LSP. It is expected that the CSP will reach 5¢/kWh by 2030[40], which may reduce the LSP's LCOE to below 3¢/kWh.

CONCLUSIONS

To conclude, here we have presented the concept of LSP, where photoluminescence allows the parallel generation of electricity from PV cells and from the residual heat that otherwise is wasted. Our experimental results show 90% of the solar photon rate can be maintained on a PV cell while the PL-absorber remains at an average temperature of 430°C. This is projected to yield electricity generation by the PV and the heat cycle having 42.5% total efficiency. In addition, LSP allows dynamic "on-demand" shift from PV power to (storable) thermal power while maintaining nearly unchanged total efficiency, which is a clear advantage over today's predetermined capacity of current solar plants. We analyzed the potential reduction in LCOE based on known CSP's LCOE and reached an average (day and night) LCOE value of 3.72¢/kWh with the potential to reach values lower than 3¢/kWh. This concept may open the way for thermal storage to become a leading solution in the increasing demand for base-load solar energy systems aiming to replace fossil fuel power plants.

EXPERIMENTAL METHODS

**Full-spectrum measurements:**

For the spectral photon rate measurements at temperatures from 25°C to 600°C, each sample is placed on a temperature-controlled heater (Thorlabs) inside an integration sphere (LabSphere 4") and excited by concentrated sunlight. The sunlight is collected by an in-lab solar concentrator system coupled to an optical fiber that is collimated and then focused onto the PL-absorber. The spectrum in the range of 300–800nm is measured by a UV-VIS spectrometer (Ocean Optics Flame-S-UV-VIS) and the spectrum in the range of 660nm– 1150nm is measured by a VIS-NIR monochromator (Andor Shamrock i303) with an InGaAs (Andor iDus) camera. At temperatures higher than 300°C, the thermal emission of the heater is measured separately and is reduced from the emission measurements. The entire setup is calibrated against a standard quartz tungsten halogen (QTH) calibration lamp (Newport). The heater's surface temperature is measured during the EQE measurements. The sample's temperature as a function of the heater's temperature is calibrated by measuring the sample's upper and lower faces temperatures as a function of the heater's temperature. The temperatures are measured by k-type thermocouples connected to a thermometer (Fluke). The presented temperatures are the average sample temperatures calculated by a calibration curve created for each sample as a function of the heater surface temperature.

**LSP efficiency simulation:**

The simulation was done using a MATLAB code developed in the lab. The standard solar spectrum for direct illumination was taken from:

**https://www.nrel.gov/grid/solar-resource/smarts.html**

## ASSOCIATED CONTENT

**Supporting Information**

The Supporting Information is available free of charge at

Additional details about experimental results of room temperature absorption coefficient (Figure s1) and EQE measurements and calculations (Figure s2); experimental procedures and results of the PV cell's performance/heat load experiment (PDF)

## AUTHOR INFORMATION

*Corresponding author. Email: carmelr@technion.ac.il

#S.H. and N.R. contributed equally to this work.

**Notes**

The authors declare no competing financial interest.

The Technion holds the following patent applications in relation to the LSP method and systems: Patent Cooperation Treaty international applications WO2014020595A2, US20150171251A1, EP2880721A2, CN104704689A (Granted 2019). The authors hold shares in Luminescent Solar—a start-up company that aims to develop this technology.

## ACKNOWLEDGMENT

The research leading to these results was supported financially by the European Union's Seventh Framework Programme (H2020/2014-2020]) under grant agreement n° 638133-ERC-ThforPV. S. Haviv acknowledges the support of the Israel Ministry of Energy. N. Kruger acknowledges the support of the Nancy and Stephen Grand Technion Energy Program (GTEP). The authors would like to thank Dr. Guy Ankonina for assisting in the absorption measurements.

TOC GRAPHICS

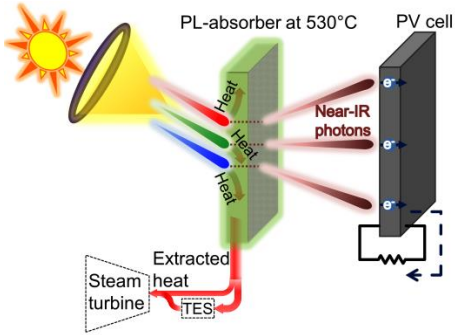